\magnification =\magstep 1
\hsize = 14.5 truecm
\vsize = 23 truecm
\overfullrule 0pt
\hoffset 1truecm
\baselineskip=24pt
\centerline{\bf Colliding Plane Waves in Einstein--Maxwell Theory}
\vskip 4truepc
\centerline{P. A. Hogan}
\centerline{Mathematical Physics Department,}
\centerline{University College Dublin,}
\centerline{Belfield, Dublin 4, Ireland}
\vskip 1truepc
\centerline{and}
\vskip 1truepc
\centerline{C. Barrab\`es and G. F. Bressange}
\centerline{Laboratoire de Math\'ematiques et Physique th\'eorique}
\centerline{UPRES-A 6083, CNRS}
\centerline{Universit\'e de Tours, 37200 France.}
\vskip 3truepc
\noindent
PACS numbers: 04.30.+x, 04.20Jb
\vskip 2truepc
Recently [1] a simple solution of the vacuum Einstein--Maxwell 
field equations was given describing a plane electromagnetic shock 
wave sharing its wave front with a plane gravitational impulse 
wave. We present here an exact solution of the vacuum Einstein--Maxwell 
field equations describing the head--on collision of such a wave 
with a plane gravitational impulse wave. The 
solution has the Penrose--Khan solution and a solution obtained 
by Griffiths as separate limiting cases.
\vfill\eject
In a recent paper[1] a construction is given of a solution of the 
vacuum Einstein--Maxwell field equations describing a plane 
electromagnetic shock wave sharing its wave front with a plane 
gravitational impulse wave. The wave is simpler than previous 
examples of such objects (see, for example [2], 
which is discussed in [1]) in that the space--time on one side of the null 
hypersurface history of the wave front is conformally flat (and 
is a special case of a Bertotti--Robinson[3] space--time) and 
on the other side is flat.The homogeneous special case of this 
wave has line--element which can be put in the form
$$ds^2=(1+b^2v^2_+)^{-1}\left\{2du\,dv-(1+lv_+)^2dx^2-
(1-lv_+)^2dy^2\right\}\ ,\eqno(1)$$
where $b, l$ are constants, $v_+=v\,\theta (v)$ with $\theta (v)$ 
the Heaviside step function (equal to 1 for $v>0$ and equal to 
zero for $v<0$). For this space--time the only non--vanishing 
Newman--Penrose component of the Maxwell field is
$$\phi _0={b\,\theta (v)\over 1+b^2v^2_+}\ ,\eqno(2)$$
and the only non--vanishing Newman--Penrose component of the 
Weyl tensor is 
$$\Psi_0=-l\,\delta (v)\ .\eqno(3)$$
Here $\delta (v)$ is the Dirac delta function. Thus both the 
Maxwell and Weyl tensors are type N in the Petrov classification 
with $\partial /\partial u$ as degenerate principal null direction. 
The null hypersurface $v=0$ is a null hyper{\it plane} and is the 
history of a plane electromagnetic shock wave on account of (2) 
and of a plane gravitational impulse wave on account of (3). We 
can remove the shock by putting $b=0$ and we can remove the 
gravitational impulse wave by putting $l=0$.
\vskip 1truepc
We consider now the head--on collision of a wave of the type 
described by (1) with a plane gravitational impulsive wave. This 
latter will be described by the space--time with line--element 
$$ds^2=2du\,dv-(1+ku_+)^2dx^2-(1-ku_+)^2dy^2\ ,\eqno(4)$$
with $k$ a constant and $u_+=u\,\theta (u)$. Following the 
usual procedure in setting up such a collision problem (see [4]) 
we consider the space--time to have line--element (1) in the 
region $u<0$ and have line--element (4) for $v<0$ (the 
two line--elements coincide in the overlapping region $u<0, v<0$). 
The line--element in the region $u>0, v>0$ (after the collision) 
has the Rosen--Szekeres form [4]
$$ds^2=2{\rm e}^{-M}du\,dv-{\rm e}^{-U}\left ({\rm e}^Vdx^2
+{\rm e}^{-V}dy^2\right )\ ,\eqno(5)$$
where $M, U, V$ are each functions of $(u, v)$ satisfying the 
O'Brien--Synge [5] junction conditions: When $v=0$
$${\rm e}^V={1+ku\over 1-ku}\ ,\qquad {\rm e}^M=1\ ,
\qquad {\rm e}^{-U}=1-k^2u^2\ ,\eqno(6)$$
and when $u=0$
$${\rm e}^V={1+lv\over 1-lv}\ ,\qquad {\rm e}^M=1+b^2v^2\ ,
\qquad {\rm e}^{-U}={1-l^2v^2\over 1+b^2v^2}\ .\eqno(7)$$
In addition the Maxwell field in the region $u>0, v>0$ has 
two non--vanishing Newman--Penrose components [4] $\phi _0, 
\phi _2$ which are both functions of $(u, v)$ and satisfy 
the boundary conditions: when $v=0, \phi _2=0$ and when $u=0, 
\phi _0=b(1+b^2v^2)^{-1}$. It is now a matter of solving the 
vacuum Einstein--Maxwell equations in the region $u>0, v>0$ 
(these can be found in [4] for example) for the unknown 
functions $U, V, M, \phi _2, \phi _0$ subject to the above 
boundary conditions. We find the following expressions for these 
functions:
$${\rm e}^{-U}={F\over 1+b^2v^2}\ ,\eqno(8)$$
$${\rm e}^V={1+ku\sqrt{1-l^2v^2}+lv\sqrt{1-k^2u^2}\over 
1-ku\sqrt{1-l^2v^2}-lv\sqrt{1-k^2u^2}}\ ,\eqno(9)$$
$${\rm e}^{-M}={H^2\over (1+b^2v^2)\left [(1-k^2u^2)(1-l^2v^2)F\right ]
^{1/2}}\ ,\eqno(10)$$
$$\phi _2={-kbv\sqrt{1-l^2v^2}\over \left [(1-k^2u^2)F\right ]^{1/2}H}\ 
,\eqno(11)$$
and
$$\phi_0={b\left\{(l^2+b^2)lkuv^3+\sqrt{1-k^2u^2}(1-l^2v^2)^{3/2}\right\}
\over (1+b^2v^2)\left [(1-l^2v^2)F\right ]^{1/2}H}\ ,\eqno(12)$$
where
$$F=1-k^2u^2-l^2v^2-k^2b^2u^2v^2\ ,\eqno(13)$$
and 
$$H=\sqrt{1-k^2u^2}\sqrt{1-l^2v^2}-kluv\ .\eqno(14)$$
A calculation of the Weyl tensor components reveals the expected 
curvature singularity at $F=0$ for $u>0, v>0$. There are two 
important special limiting cases: (1) if $b=0$ the solution above 
becomes the Penrose--Khan [6] solution describing the space--time 
following the collision of two plane impulsive gravitational waves 
and (2) if $l=0$ the solution becomes the Griffiths [4, 7] solution 
describing the space--time following the collision of a plane 
gravitational impulse wave and a plane electromagnetic shock wave. 
Clearly further collisions involving the type of plane wave described 
here by (1) can be envisaged.
\vskip 1truepc
\noindent
We thank Professor Werner Israel for helpful and encouraging 
discussions.
\vskip 4truepc
\noindent
{\bf References:}
\vskip 1truepc
\item{[1]} Hogan, P. A. 1997 ``Plane Gravitational Waves and 
Bertotti--Robinson Space--Times'', University College Dublin, preprint.
\vskip 1truepc
\item{[2]} Chandrasekhar, S. and Xanthopoulos, B. C. 1985 {\it Proc. 
R. Soc.} A{\bf 398}, 223.
\vskip 1truepc
\item{[3]} Kramer, D., Stephani, H., Mac Callum, M. A. H. and 
Herlt, E. {\it Exact Solutions of Einstein's Equations} (VEB Deutscher 
Verlag der Wissenschaften, Berlin 1980).
\vskip 1truepc
\item{[4]} Griffiths, J. B. {\it Colliding Plane Waves in General 
Relativity} (Clarendon Press, Oxford 1991).
\vskip 1truepc
\item{[5]} O'Brien, S. and Synge, J. L. 1952 {\it Commun. Dublin 
Inst. Adv. Stud.} A, no.9.
\vskip 1truepc
\item{[6]} Penrose, R. and Khan, K. A. 1971 {\it Nature} {\bf 229}, 185.
\vskip 1truepc
\item{[7]} Griffiths, J. B. 1975 {\it Phys. Lett.}, A{\bf 54}, 269.

\bye